\documentclass{jetp}
\usepackage{setspace}
\usepackage{graphicx}

\begin{document}
%\onehalfspacing
%\onehalfspacing
\English
\title{Dark matter constraints  from an observation of
  dSphs and the LMC with the Baikal NT200} 

%\rtitle{Constraint in dark matter signal from \ldots
%dSphs and the LMC with the Baikal NT200\ldots}

%%% article title - for table of contents (usualy identical with \title)
%\sodtitle{}

\setauthor{A.~D.}{Avrorin}{1}
\setauthor{A.~V.}{Avrorin}{1}
\setauthor{V.~M.}{Aynutdinov}{1}
\setauthor{R.}{Bannasch}{9}
\setauthor{I.~A.}{Belolaptikov}{2}
\setauthor{V.~B.}{Brudanin}{2}
\setauthor{N.~M.}{Budnev}{3}
\setauthor{I.~A.}{Danilchenko}{1}
\setauthor{S.~V.}{Demidov}{1}\email{demidov@ms2.inr.ac.ru}
\setauthor{G.~V.}{Domogatsky}{1}
\setauthor{A.~A.}{Doroshenko}{1}
\setauthor{R.}{Dvornicky}{27}
\setauthor{A.~N.}{Dyachok}{3}
\setauthor{Zh.-A.~M.}{Dzhilkibaev}{1}
\setauthor{L.}{Fajt}{78}
\setauthor{S.~V.}{Fialkovsky}{5}
\setauthor{A.~R.}{Gafarov}{3}
\setauthor{O.~N.}{Gaponenko}{1}
\setauthor{K.~V.}{Golubkov}{1}
\setauthor{T.~I.}{Gress}{3}
\setauthor{Z.}{Honz}{2}
\setauthor{K.~G.}{Kebkal}{9}
\setauthor{O.~G.}{Kebkal}{9}
\setauthor{K.~V.}{Konischev}{2}
\setauthor{A.~V.}{Korobchenko}{3}
\setauthor{A.~P.}{Koshechkin}{1}
\setauthor{F.~K.}{Koshel}{1}
\setauthor{A.~V.}{Kozhin}{4}
\setauthor{V.~F.}{Kulepov}{5}
\setauthor{D.~A.}{Kuleshov}{1}
\setauthor{M.~B.}{Milenin}{5}
\setauthor{R.~A.}{Mirgazov}{3}
\setauthor{E.~R.}{Osipova}{4}
\setauthor{A.~I.}{Panfilov}{1}
\setauthor{L.~V.}{Pan'kov}{3}
\setauthor{E.~N.}{Pliskovsky}{2}
\setauthor{M.~I.}{Rozanov}{6}
\setauthor{E.~V.}{Rjabov}{3}
\setauthor{B.~A.}{Shaybonov}{2}
\setauthor{A.~A.}{Sheifler}{1}
\setauthor{M.~D.}{Shelepov}{1}
\setauthor{A.~V.}{Skurihin}{4}
\setauthor{O.~V.}{Suvorova}{1}\email{suvorova@cpc.inr.ac.ru}
\setauthor{V.~A.}{Tabolenko}{3}
\setauthor{B.~A.}{Tarashansky}{3}
\setauthor{S.~A.}{Yakovlev}{9}
\setauthor{A.~V.}{Zagorodnikov}{3}
\setauthor{V.~L.}{Zurbanov}{3}
          
\setaffiliation1{Institute for Nuclear Research RAS, 117312 Moscow,
  Russia}
\setaffiliation2{Joint Institute for Nuclear Research, Dubna,
  Russia}
\setaffiliation3{Irkutsk State University, Irkutsk, Russia}
\setaffiliation4{Skobeltsyn Institute of Nuclear Physics MSU,
  Moscow, Russia}
\setaffiliation5{Nizhni Novgorod State Technical University, Nizhni
  Novgorod, Russia}
\setaffiliation6{St.Petersburg State Marine University,
  St.Petersburg, Russia}
\setaffiliation7{Comenius  University, Mlynsk\'a  dolina  F1,  SK-842  48  Bratislava, Slovakia}
\setaffiliation8{Czech  Technical  University  in  Prague,  12800 Prague, Czech  Republic}
\setaffiliation9{EvoLogics GmbH, Berlin, Germany}

%%% abstract
\abstract{
%\begin{onehalfspace}
In present analysis we complete search for a dark matter signal
with the Baikal neutrino telescope NT200 from potential sources in
the sky.  We use five years of data  and look for neutrinos from dark matter
  annihilations in  the dwarfs spheroidal galaxies in the Southern hemisphere
and the Large Magellanic Cloud known as the largest and close satellite galaxy 
of the Milky Way. We do not find any excess in observed data over
expected background from the atmospheric neutrinos towards the LMC or
any of tested 22 dwarfs. We perform a joint likelihood analysis on the
sample of five selected dwarfs and found a concordance of the data
with null hypothesis of the background-only observation.
We derive 90\% CL upper limits on the cross section of annihilating
dark matter particles of mass between 30 GeV and 10 TeV into several
channels both in our combined analysis of the dwarfs and in a
  particular analysis towards the LMC.

%\end{onehalfspace}
}
%%% PACS numbers
%\PACS{74.50.+r, 74.80.Fp}

%\begin{document}

\maketitle
\begin{onehalfspace}

\section {Introduction}

Vast majority of astrophysical and cosmological observational data
indicate on existence of new particles -- dark
matter (DM)~\cite{Bertone:2004pz}. 
The natural and the most favorable candidate for this phenomena 
is Weakly Interacting Massive Particles
(WIMP)~\cite{Steigman:1984ac}. Their main property is possibility of
annihilation into pairs of ordinary matter particles followed by their
decays and hadronization. Searches for such annihilation signal from
different astrophysical objects are performed by numerous experiments
including neutrino telescopes.  

The dwarf spheroidal galaxies (dSphs) being distant satellites of the
Milky Way (MW) are optically faint or ultra-faint astrophysical
objects, with the visible angular sizes less than about degree
appeared to an observer. The dwarfs are characterized by a very
small light-to-mass 
ratios and  the flat rotation curves (see
e.g. Refs.~\cite{Strigari2008na} and references therein),  
that suggests considerable dark matter content. As a consequence, the
dSphs are considered as sources of significant dark matter annihilation
signal, while the experimental controversies are widely discussed (see
e.g. Refs.\cite{Klypin2004apj, Kravtsov2010aa, Hammer2014mnra,
  Klypin2016mnra}). Upto now about four dozens of dSphs have been
discovered, with account of new eight dwarfs recently 
found by the DES~\cite{Bechtol:2015cbp,Drlica-Wagner:2015ufc}. Their
visible population in the sky might be increased significantly due to
new optical imaging surveys i.e. DES~\cite{DES-2005} and
LSST~\cite{LSST-2009} and complementary observations with the
gamma-telescopes of a next generation like the CTA~\cite{cta-2012ap}. 
This will open new opportunities in the DM investigation. Presently, among 
many multimessenger searches for DM annihilation signal from
extragalactic sources like the dSphs, there was not found significant
excess in the number of events relatively to estimated background.
Furthermore, there are robust upper bounds on the DM annihilation
signal in the joint analysis of the dSphs obtained with the
gamma-telescopes FERMI-LAT and
DES~\cite{Ackermann:2015zua,Drlica-Wagner:2015xua,Fermi-LAT:2016uux},
also the MAGIC~\cite{Aleksic:2013xea} and the
HESS~\cite{::2016jja}. Similar search of the indirect DM signal has
been performed with the IceCube~\cite{Aartsen:2015bwa} and as 
well in the current work on base of the neutrinos detection. Evidence
of dark matter repository in the nearby and the largest satellite
galaxy of the Milky Way i.e. the Large Magellanic Cloud (LMC)  makes this region by the second
brightest astrophysical source of DM annihilations~\cite{LMC-2004ap}
 after the MW Galactic Center (GC)~\cite{GCE2012phr,
   fermilat2015arx}. 
The recent analysis of the FERMI-LAT data~\cite{Buckley:2015doa} shows
that the obtained upper limits 
on the DM annihilation signal in the LMC are quite competitive with
those from the dSphs, despite the LMC unlike the dwarfs has
significant baryonic background.

High energy neutrinos incoming from the directions towards the dwarfs
or the LMC are expected to be among the copious particles generated in
the different DM annihilation channels. If these neutrinos have their
energies above an energy threshold, they have probability to be
detected by neutrino telescopes. A major challenge is to suppress
background of the atmospheric muons exceeding upward going
neutrino flux by the factor $10^6$. A particular aim is to resolve
non-atmospheric origin of neutrinos towards the DM sources in a search
for indirect signal of dark matter taking into account known
uncertainties both experimental, theoretical and astrophysical.   

In the present analysis we use the NT200 neutrino dataset, perform a
search for an excess in the directions towards the LMC and dSphs and
put constraints on dark matter annihilation cross section.
The paper is organized as follows. In section 2 we shortly
recall the experiment NT200 and the data selection. In section 3 we
describe signal and background properties and their simulation for the
the cases of dwarfs galaxies and the LMC.  In section 4 we obtain
upper limits on dark matter annihilation cross section and make a
comparison with the results from other experiments. Conclusions are
presented in section 5.  

\section {Experiment and observation}
In a search for dark matter signal, we use measured directions of muons
arriving to the Baikal neutrino telescope NT200. This is indirect
method to estimate a signal strength with subsequent off-line
optimization of signal to 
background ratio. Cherenkov radiation of muons and hadronic
showers induced by relativistic neutrino scatterings off nucleons in
surrounding water is distinctive by fixed Cherenkov angle (about
42$^\circ$ in water) relatively the track of moving relativistic
particle at each point of its path. It is a base in reconstruction of
arrival time of particle and its angular coordinates. The Cherenkov
emission is effectively collected by optical modules (OMs) in
3D-configuration of photodetectors on strings, while the efficiency of
this detection is strongly determined by hydro-optically properties of
medium (see e.g. Ref.~\cite{UFN-2015} and references therein). 
%%%
%%%    Ris_1
%%%
\begin{figure}
\begin{center}
\includegraphics[width=0.33\textwidth,angle=0]{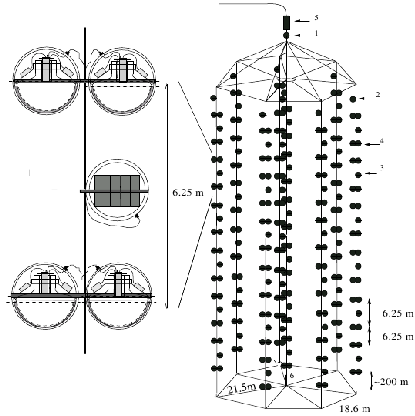}
 \end{center}
\caption{\label{fig:BaikalNT200} 
NT200 schematic view: 1 -- array electronics module; 2 -- string
electronics module; 3 -- ''svjaska'' (shown on left separately)
i.e. two pairs of OMs with electronics module;  4 --  pair of OMs; 5
and 6 -- calibration lasers.}  
\end{figure}

The telescope NT200 is placed at 51.83$^\circ$ North latitude in the
southern basin of the lake Baikal, at a distance of 3.5 km off the
shore and at a depth of \mbox{1.1 km}, its instrumentation volume
encloses 100 Ktons. The optical properties of the 
Baikal water are characterized by the absorption length $20\div24$ m
at 480 nm and the scattering length $30\div70$ m depending on a
season. Recall (see Fig.~\ref{fig:BaikalNT200}) that the detector
NT200 consists of 192 optical modules arranged in pairwise at twelve
storeys (''svjaska'' is shown with zoom) on 8 strings of 72\,m length:
seven peripheral strings and a central one. The distances between the
strings are about 21.5\,m. They are placed at vertices of heptagon
with side of 18.5\,m of size. Each OM contains hybrid photodetector
QUASAR-370, a photo multiplier tube (PMT) with 37-cm diameter. To
suppress background from dark noise, the two PMTs of a pair are
switched in coincidence within a time window of 15\,ns. The OMs are
time-synchronized and energy-calibrated by artificial light
pulses. The operated configurations of the NT200, its functional
systems and the calibration methods have been described
elsewhere~\cite{NT200-1997,NT200-2006,Baikal2007,Baikal2009}. 

\begin{figure}[!htb]
\begin{center}
\includegraphics[width=0.33\textwidth,angle=0]{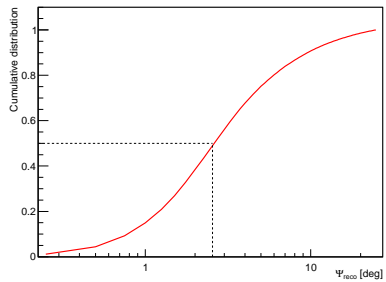}
 \end{center}
\caption{\label{fig:PsiReco} 
The cumulative distribution of the mismatch angle $\Psi_{reco}$ in the
NT200 dataset reconstruction. The black-dotted line indicates the
median value.}   
\end{figure}

\begin{figure}[!htb]
\begin{center}
\includegraphics[width=0.33\textwidth,angle=0]{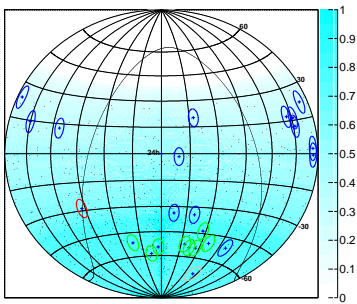}
 \end{center}
\caption{\label{fig:skyVis+22dSph} 
A skymap of time visibility for the Baikal NT200 and data sample
in equatorial coordinates with placed 14 classical dSphs (blue
circles), 8 new discovered faint dwarfs (yellow), the Galactic Center
(red) and the LMC (pink).}   
\end{figure}

In present analysis of the Baikal NT200 survey of the sky we used
dataset for five years between April of 1998 and February of 2003,
with in total 2.76 live years. The same dataset and the same Monte
Carlo (MC) sample has been implemented in our searches for a DM
annihilation signal towards the Sun~\cite{NT200Sun:2014swy} and the
Galactic Center~\cite{NT200GC:2015}. In particular, we have used
dataset selected by the off-line filter with requirements of at least
6 hits on at least 3 strings ("muon trigger 6/3"), that selects about
$40\%$ of all triggered events with angular resolution of about
$14.1^\circ$ in term of the r.m.s. mismatch angle $\psi_{reco}$. To
get the best possible estimator for the direction, we use multiple
start guesses for the $\chi^{2}$
minimization~\cite{NT200Astro-2011}. Basic analysis on distinguish
upward and downward going muons on a one-per-million mis-assignment
level has been done earlier~\cite{Belolap07}, where the code was
developed for the atmospheric neutrinos ($\nu_{atm}$). Study of the
NT200 response on the fluxes of atmospheric muons and neutrinos has
been done with the MC simulations based on the standard codes
{\texttt{CORSIKA}}~\cite{CORSIKA} and {\texttt{MUM }}\cite{MUM} using
the Bartol atmospheric $\nu$ flux~\cite{Bartol}. For the final filter
of events the quality parameters are applied and they are not related
to the time information e.g. variables like the number of hit
channels, $\chi^2/d.o.f.$, the probability of fired channels to have
been hit and not fired channels haven't been hit and the actual
position of the track with respect to 
the detector center (for more details see
Ref.~\cite{NT200Astro-2011,Belolap07}). To suppress background muons
which are downgoing near horizon we select only the events with
the 
reconstructed zenith angle $\Theta > 100^\circ$. All the cuts provide
rejection factor for the atmospheric muons of about $10^{-7}$.
Selection by the quality criteria results in data sample of 510 events
for current analysis, having the neutrino energy threshold of about
10~GeV and median value $2.5^\circ$ in the mismatch angles
distribution, as it is seen in Fig.~\ref{fig:PsiReco}.
%%%
%%%    Ris_2
%%%
In MC simulations of signal events to be described in the next section
we have applied the shape of differential $\Psi_{reco}$
distribution. The distribution of arrival directions of selected
events is presented in Fig.\ref{fig:skyVis+22dSph}
%%%
%%%    Ris_2
%%%
by black points along background of sky visibility, i.e. the part of
time during which the particular direction is observed by NT200, shown
by gradient in color. Due to the NT200 location and the selection of
neutrino events from the lower hemisphere only, in what follows we
consider the sources of dark matter annihilation signal with
declinations below 39$^\circ$N. Full visibility is reached for
declinations larger 39$^\circ$S.

\section{Signal and background}
In search for neutrino fluxes above the expected background of atmospheric
neutrinos, we test 23 directions including those towards a set of
dSphs and the LMC on the skymap. In Fig.\ref{fig:skyVis+22dSph} are
drawn the coordinates of 14 classical dark dwarf galaxies by blue
crosses (and 5 degrees circles around them): "Carina", "Fornax",
"Leo-I","Leo-II", "Sculptor", "Sextans", "Bootes-I", "Coma Berenices", 
"Hercules", "Leo-IV", "Leo-V", "Leo-T", "Segue-1", "Segue-2". There
are also 8 ultra faint dSphs discovered recently by the DES
collaboration~\cite{Bechtol:2015cbp} at declinations larger
45$^\circ$S. They are marked by yellow
5 degrees circles in Fig.\ref{fig:skyVis+22dSph}: "Reticulum-2",
"Eridanus-2", "Horologium-1", "Pictor-1", "Phoenix-2", "Indus-1",
"Eridanus-3", "Tucana-2". In Table~\ref{dwarfs_pars} all 22 dwarf
galaxies are presented along with their galactic coordinates
($\delta$, $\alpha$).  
\begin{table}[!htb]
  \begin{center}
    \begin{tabular}{|c|c|c|c|c|c|c|c|}
      \hline
      Name & Dec & RA & $\overline{\log_{10}{J}}$ & $N_{S}$ & $N_{B}$ & TS, $b\bar{b}$ 30~GeV & TS, $\nu\bar{\nu}$ 10~TeV \\\hline
      Carina & -50.97 & 100.40 & 18.1$\pm0.23^a$ & 30 & 29.3 & 0.10 & 1.11 \\
      Fornax & -34.45 & 40.0 & 18.2$\pm0.21^a$ & 25 & 26.0 & 0.02 & 0 \\ 
      Leo-I & 12.31 & 152.12 & 17.7$\pm0.18^a$ & 14 & 11.6 &1.15 &0.05  \\ 
      Leo-II & 22.15 & 168.37 & 17.6$\pm0.18^a$ & 11 & 6.81 & 2.19 & 0  \\
      Sculptor & -33.71 & 15.04 & 18.6$\pm0.18^a$ & 29 & 24.5 & 0 & 0.25 \\ 
      Sextans & -1.61 & 18.26 & 18.4$\pm0.27^a$ & 23 & 17.6 & 2.50 & 0 \\ 
      Bootes-I & 14.50 & 210.03 & 18.8$\pm0.22^a$ & 12 & 10.7 & 0.05 & 0.95 \\ 
      Coma Berenices & 23.90 & 186.75 & 19.0$\pm0.25^a$ & 10 & 6.12 & 0.76 & 0.12 \\
      Hercules & 12.79 & 247.76 & 18.1$\pm0.25^a$ & 9 & 11.3 & 0 & 0 \\
      Leo-IV & -0.53 & 173.24 & 17.9$\pm0.28^a$ & 18 & 16.8 & 0.0 & 0.48 \\
      Leo-V & 2.22 & 172.79 & 16.37$\pm0.9^b$ & 18 & 15.4 & 0.0 & 0 \\
      Leo-T & 17.05 & 143.72 & 17.11$\pm0.4^b$ & 14 & 9.34 & 0 & 0 \\
      Segue-1 & 16.08 & 151.77 & 19.5$\pm0.29^a$ & 13 & 9.76 & 1.28 & 0.78 \\
      Segue-2 & 20.18 & 34.82 & 16.21$\pm1.0^b$ & 8 & 7.83 & 0.03 & 0.99 \\
      Reticulum-2 & -54.05 & 53.92 & 19.8$\pm0.9^c$ & 20 & 28.7 & 0.01 & 0.76 \\
      Eridanus-2 & -43.53 & 56.09 & 17.3$\pm0.4^d$ & 25 & 27.5 & 0 & 0 \\ 
      Horologium-1 & -54.11 & 43.87 & 18.4$\pm0.4^d$ & 22 & 28.8 & 1.02 & 0 \\ 
      Pictor-1 & -50.28 & 70.95 & 18.1$\pm0.4^d$ & 19 & 28.6 & 0 & 0 \\ 
      Phoenix-2 & -54.41 & 354.99 & 18.4$\pm0.4^d$ & 35 & 28.2 & 2.34 & 0 \\ 
      Indus-1 & -51.16 & 317.20 & 18.3$\pm0.4^d$ & 28 & 27.3 & 0 & 0 \\ 
      Eridanus-3 & -52.28 & 35.69 & 18.3$\pm0.4^d$ & 29 & 28.7 & 0.63 & 4.96 \\ 
      Tucana-2 & -58.57 & 343.06 & 18.8$\pm0.4^d$ & 31 & 27.4 & 2.38 & 1.98 \\\hline 
    \end{tabular}
  \end{center}
  \caption{\label{dwarfs_pars} List of dwarf galaxies used in the
    present analyses, their coordinates and $J_a$-factors integrated
    over 20$^\circ$ opening angle, number of observed events
    $n$ and expected number of background events $N_B$ in the
    signal region, test statistics (TS) for supposed signal with
    $m_{DM}=30$~GeV and $b\bar{b}$ channel and with $m_{DM}=10$~TeV
    and $\nu\bar{\nu}$ channel. Uncertainties in the $J_a$-factors are
    marked with superscripts corresponding to the following
    references: $a$)~\cite{Ackermann:2013yva},
    $b$)~\cite{Geringer-Sameth:2014yza},
    $c$)~\cite{Bonnivard:2015tta},
    $d$)~\cite{Drlica-Wagner:2015xua}. }  
\end{table}

Neutrino flux from dark matter annihilations in a galaxy in direction
at an angle $\psi$ with respect to its center is given by 
\begin{equation}
\label{eq:1}
\frac{d\phi_{\nu}}{dE_{\nu}d\Omega} = J_{a}(\psi)\;\frac{\langle
  \sigma_a v\rangle}{8\pi m_{DM}^2}\;\frac{dN_{\nu}}{dE_{\nu}}.
\end{equation}
Part of this expression depends on particle physics properties of dark
matter: $\langle\sigma_av\rangle$ is its annihilation cross section in
present time, $dN_\nu/dE_\nu$ is energy spectrum of neutrinos
generated in decays processes of products of DM annihilations in  
dependence on masses of dark matter particles $m_{DM}$. In what
follows we consider the DM mass interval from 30~GeV to 10~TeV.
Neutrino energy spectra depend on properties of dark matter
annihilation. We consider a set of five benchmark annihilation 
channels: $b\bar{b}$, $\tau^+\tau^-$, $\mu^+\mu^-$, $W^+W^-$ and
$\nu\bar{\nu}\equiv \frac{1}{3}(\nu_e\bar{\nu}_e +
\nu_\mu\bar{\nu}_\mu + \nu_\tau\bar{\nu}_\tau)$. For the present
analysis the neutrino energy spectra in these channels have been  
taken from Ref.~\cite{Baratella:2013fya}. Due to the loss of coherence
after propagation over cosmologically large distances neutrino arrive
at the Earth as mass states and to calculate $\nu_\mu$
($\bar{\nu}_\mu$) energy spectra we use the following 
set~\cite{Forero:2014bxa} of neutrino oscillation parameters:
$\Delta m_{21}^2 = 7.6\cdot 10^{-5}~{\rm eV}^2$, $\Delta m_{31}^{2} =
2.48\cdot 10^{-3}~{\rm eV}^2$, $\delta_{CP} = 0$, $\sin^2{\theta_{12}}
= 0.323$, $\sin^2{\theta_{23}} = 0.567$, $\sin^2{\theta_{13}} = 0.0234$.

Astrophysical part of the neutrino flux~\eqref{eq:1} is encoded in the
values of $J_a$-factors which depend on the dark matter content
  and angular distance between
direction of observation and the direction towards the center of the
corresponding source. This quantity is given by the following integral
over line-of-sight 
\begin{equation}
\label{eq:2}
J_a(\psi) = \int_{0}^{l_{max}}dl\;
\rho^2\left(\sqrt{R_0^2-2lR_0\cos{\psi}+l^2}\right)
\end{equation}
of dark matter density $\rho(r)$ squared. In eq.~\eqref{eq:2}, $R_0$
is the distance from the center of the source to the Solar System.
The larger value of annihilation factor $J_a(\psi)$, the larger
magnitude of the neutrino flux is expected at the Earth. Total
expected number of signal events to be detected by neutrino telescope
NT200 from dark matter annihilations in a distant source can be
calculated using eq.~\eqref{eq:1} as follows  
\begin{equation}
  \label{eq:3}
  N_S = T\frac{\langle\sigma_a v\rangle}{8\pi m_{DM}} J_{\Delta\Omega}
  \int_{E_{\th}}^{m_{DM}}dE_\nu\frac{dN_\nu}{dE_\nu}S_\nu(E_\nu).
\end{equation}
Here $T$ is the livetime, $S_\nu(E_\nu)$ is the effective area of the
telescope for neutrinos coming from the direction towards the source
in quest and
\begin{equation}
  \label{eq:3_1}
J_{\Delta\Omega} = \int d(\cos{\psi})d\phi J(\psi),
\end{equation}
where the integral is taken over a search region to be discussed below. 
The effective area $S_\nu(E_\nu)$ is calculated from MC simulations
(see section 1) by determination of detection efficiency of muon neutrino 
coming from a particular direction. We refer reader to
Refs.~\cite{NT200Sun:2014swy,NT200GC:2015} for detailed discussions.

In the following analysis we consider the distant dwarfs galaxies as
point like sources since theirs angular sizes are well within angular
resolution of the NT200. In Table~\ref{dwarfs_pars} we present the
values of $J$-factor~\eqref{eq:3_1} for these galaxies integrated over  
solid angle corresponding to their sizes. Also we include the
astrophysical uncertainties in the estimates of $J$-factors with
corresponding references. In the chosen set of dwarf galaxies the
largest $J_a$-factor is expected from ``Reticulum-2''. As it has been
discussed in Ref.~\cite{Bonnivard:2015tta}, this source is a very
attractive target in search for DM signal.

Great interest to test direction towards the LMC is motivated by
recent dark matter analysis of FERMI-LAT data performed
in~\cite{Buckley:2015doa} indicating the LMC to be the next after
  the GC bright source of the annihilation
signal. Position of this source on the sky allows for 100\% time
observation in neutrinos by NT200. In performance of the DM signal
from the LMC we take into account extended size of this galaxy and
simulate neutrino signal using dark matter density
profile~\cite{Hernquist:1990be,Zhao:1995cp,Kravtsov:1997dp} of the
following functional form   
\[
\rho(r) = \frac{\rho_0}{\left(\frac{r}{r_S}\right)^\gamma\left[1 +
    \left(\frac{r}{r_S}\right)^\alpha\right]^{\frac{\beta-\gamma}{\alpha}}}
\theta(r_{max}-r).
\]
Here $r_{max}=100$~kpc, $\theta(r)$ is the step function.
Following Ref.~\cite{Buckley:2015doa} we consider three dark matter
halo profiles to be referred as {\it sim-mean, sim-min} and {\it
  sim-max}.  Corresponding
parameters $\alpha$, $\beta$, $\gamma$ and $r_S$ are presented in
Table~\ref{LMC_pars}.
\begin{table}[!htb]
  \begin{center}
    \begin{tabular}{|c|c|c|c|c|c|c|}
    \hline
    Profile & $\alpha$ & $\beta$ & $\gamma$ & $r_S$, kpc & $\rho_0$,
    GeV$/$cm$^3$ & $\overline{\log_{10}{J}}$ \\\hline
    {\it sim-max} & 0.35 & 3.0 & 1.3 & 5.4 & 4.19 & 21.94\\\hline
    {\it sim-mean} & 0.96 & 2.85 & 1.05 & 7.2 & 0.32 & 20.38\\\hline
    {\it sim-min} & 1.56 & 2.69 & 0.79 & 4.9 & 0.46 & 20.25\\\hline
  \end{tabular}
  \caption{\label{LMC_pars} Parameters of dark matter halo profiles for
  Large Magellanic Cloud~~\cite{Buckley:2015doa}.}
\end{center}
\end{table}
We used {\it sim-mean} profile to obtain the upper limits and the
others, i.e. {\it sim-min} and {\it sim-max}, to estimate the influence of
astrophysical systematic uncertainties. We note that there is also an 
uncertainty in determination of the gravitational center of the
LMC. Its size is within about 1.5$^\circ$~\cite{Buckley:2015doa}
which is considerably smaller than the angular resolution of NT200. In 
the present analysis we choose the coordinates of the LMC center
$l=280.54^\circ, b=-32.51^\circ$ derived from stellar rotation
curves~\cite{vanderMarel:2002kq}. We see that $J_a$-factor of LMC
integrated over search region is larger than that of all dwarfs
(see Tables~\ref{dwarfs_pars} 
and~\ref{LMC_pars}) which  along with 100\% visibility
makes this source very attractive for searches for dark matter
signal. 

\begin{figure}
\begin{center}
\includegraphics[width=0.33\textwidth,angle=-90]{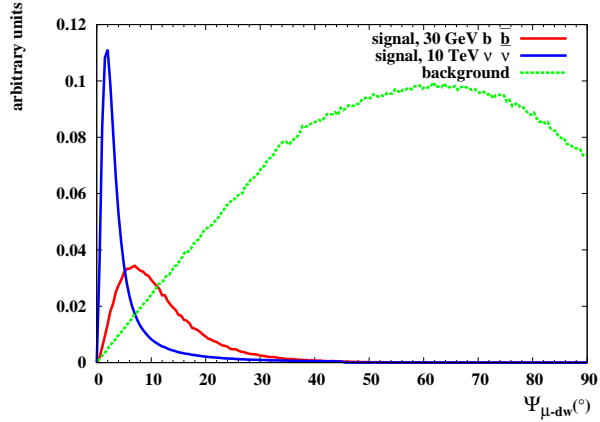}
 \end{center}
\caption{\label{fig:sig_back} 
  Angular distributions of signal ($b\bar{b}$, $m_{DM}=30$~GeV and
  $\nu\bar{\nu}$, $m_{DM}=10~TeV$) and background events for
  ``Reticulum-2'' as a source. }    
\end{figure}

We simulate the expected energy and angular distribution of the signal
events from dwarf galaxies and LMC as described in
Ref.~\cite{NT200GC:2015}. Atmospheric neutrinos are dominating
source of the background. In the present analysis we estimate the
expected background from the data using scrambling by randomization of
RAs of the observed events. This procedure can not exclude a possible
signal contamination in our determination of the background which is
however expected to be small. In Fig.~\ref{fig:sig_back} we present comparison
of angular distribution of signal and background for ``Reticulum-2''
for an example. Here we show angular distributions of the signal for the softest ($b\bar{b}$,
$m_{DM}=30$~GeV) and hardest ($\nu\bar{\nu}$, $m_{DM}=10$~TeV) 
  among the chosen
annihilation channels.  
We see that the angular spread of the signal varies up to
$20-25^\circ$; the most collimated signal is produced in
monochromatic neutrino channel.

\section{Data analysis}
The analysis is performed by construction of likelihood function 
  along the same lines as described in Ref.~\cite{NT200GC:2015}. For
obtained angular 
spreads of signal events from the sources we choose as search region
a cone with half-cone angle of 20$^\circ$ around each single source.
The numbers of observed and background events within this cone 
for each probing direction towards corresponding dwarf galaxy are
presented in Table~\ref{dwarfs_pars}. In the search region of the LMC 
the number of observed events is 23 and the expected background is
estimated as 29.5.

Given the expected angular signal and background distributions for a
source, $f_S(\psi)$ and $f_B(\psi)$, respectively, the resulting
angular distribution has the form 
\begin{equation}
f(\psi,N_S,N_B) = \frac{1}{N_S+N_B}\left(N_Sf_S(\psi) + N_Bf_B(\psi)\right),
\end{equation}
where $N_S$ and $N_B$ are expected number of signal and background
events inside the search region, respectively. The likelihood function
for a particular source can be written as follows
\begin{equation}
  \label{eq:4}
  {\cal L}(\langle\sigma_a v\rangle) =
  \frac{(N_B+N_S)^n}{n!}{\rm e}^{-(N_B+N_S)}\times \prod_{i=1}^{n}
  f(\psi_i,N_B,N_S), 
\end{equation}
where $n$ is the observed number of events and the first multiplier
accounts for statistical fluctuations in $n$. To take into account
systematic 
uncertainties the likelihood function is modified by introducing a set
of nuisance parameters $\theta=\{\epsilon_S,\epsilon_B,J\}$ which
change~\eqref{eq:4} as
\begin{equation}
  {\cal L}(\langle\sigma_a v\rangle,\theta) = {\cal N}
  \frac{(\epsilon_BN_B+\epsilon_SN_S)^n}{n!}
  {\rm e}^{-(\epsilon_BN_B + \epsilon_SN_S) -
  \frac{(\epsilon_S-1)^2}{2\sigma_S^2} -
  \frac{(\epsilon_B-1)^2}{2\sigma_B^2} -
  \frac{({\rm log}_{10}(J) - \overline{{\rm
        log}_{10}(J)})^2}{2\sigma_J^2}}
  \prod_{i=1}^{n}   f(\psi_i,\epsilon_BN_B,\epsilon_SN_S).
\end{equation}
Here the dependence on $\langle\sigma_a v\rangle$ and $J$ enters
implicitly through $N_S$ via Eq.~\eqref{eq:3}. In the above expression
$\sigma_S$ and $\sigma_J$ are particle physics and astrophysical 
systematic in the signal, while $\sigma_B$ is the systematic
uncertainty in the background. We intentionally divide systematic
uncertainties in number of signal events coming from particle physics
and from astrophysics. The latter enters into the likelihood function
through estimates of error in $J_a$-factors presented
in Table~\ref{dwarfs_pars}. For careful discussion of other sources of
systematic uncertainties we refer reader to
Refs.~\cite{NT200Sun:2014swy,NT200GC:2015}. Their size is dominated by
30\% experimental uncertainty resulting from optical properties of
water and in the sensitivity of the optical modules. Theoretical
errors reach 10-12\% coming from uncertainties in neutrino oscillation
parameters and neutrino-nucleus interaction cross section.

The upper limits on $\langle\sigma_a v\rangle$ are then obtained using
profile likelihood  constructed as
\begin{equation}
\label{eq:5}
  \lambda(\langle\sigma_a v\rangle) = -2\;{\rm ln}\frac{{\cal L}(\langle\sigma_a v\rangle,
  \hat{\hat{\theta}}(\langle\sigma_a v\rangle))}{{\cal L}(\widehat{\langle\sigma_a v\rangle}, 
  \hat{\theta})}.
\end{equation}
Here  $\widehat{\langle\sigma_a v\rangle}$ and $\hat{\theta}$ are the
values which give absolute maximum to the likelihood probability
function in physical region with $\langle\sigma_a v\rangle \geq 0$, while $\hat{\hat{\theta}}(\langle\sigma_a v\rangle)$ denotes
the value of the nuisance parameters $\theta$ in the maximum of the
likelihood at fixed value of $\langle\sigma_a v\rangle$. For analysis
with the LMC we do not include astrophysical systematic in the 
likelihood  function but instead we calculate upper limits for
different 
dark matter density profiles {\it sim-min} and {\it sim-max} 
  considering them as limiting cases (see discussion of sources of
  astrophysical uncertainties in~\cite{Buckley:2015doa}). In the
joint analysis with several dwarf galaxies corresponding likelihood
function is constructed as a product of individual likelihood
functions as follows 
\begin{equation}
\label{eq:6}
  {\cal L}(\langle\sigma_a v\rangle,\theta) = \prod_{i=1}^{N_d}{\cal
    L}_i(\langle\sigma_a v\rangle,\theta_i).
\end{equation}
Here $N_d$ is the number of sources taken in the joint analysis and
${\cal L}_i(\langle\sigma_a v\rangle,\theta_i)$ is the individual
likelihood function of $i$-th source.

\begin{figure}
  \begin{center}
    \begin{tabular}{cc}
      \includegraphics[width=0.3\textwidth,angle=-90]{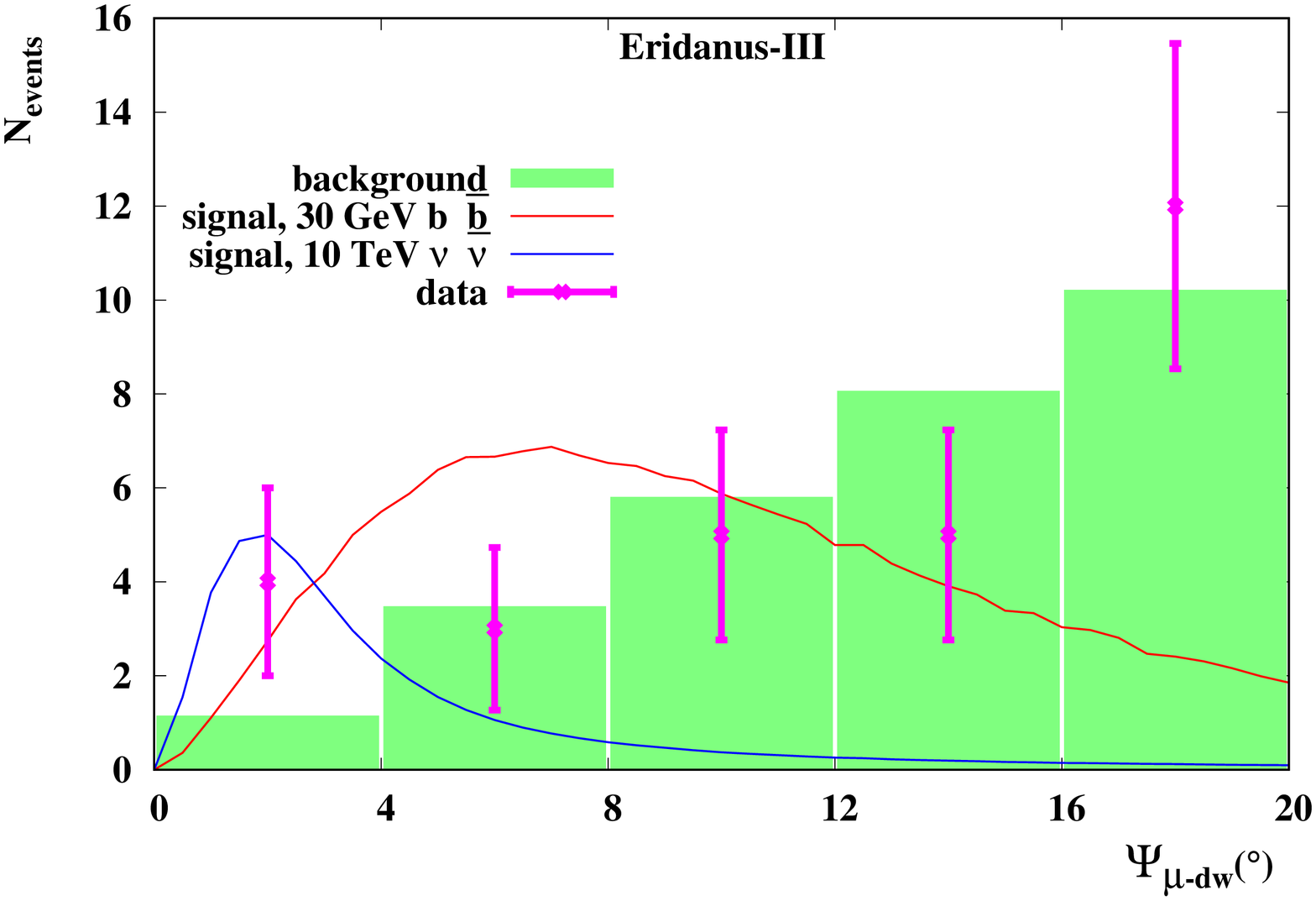} &
      \includegraphics[width=0.3\textwidth,angle=-90]{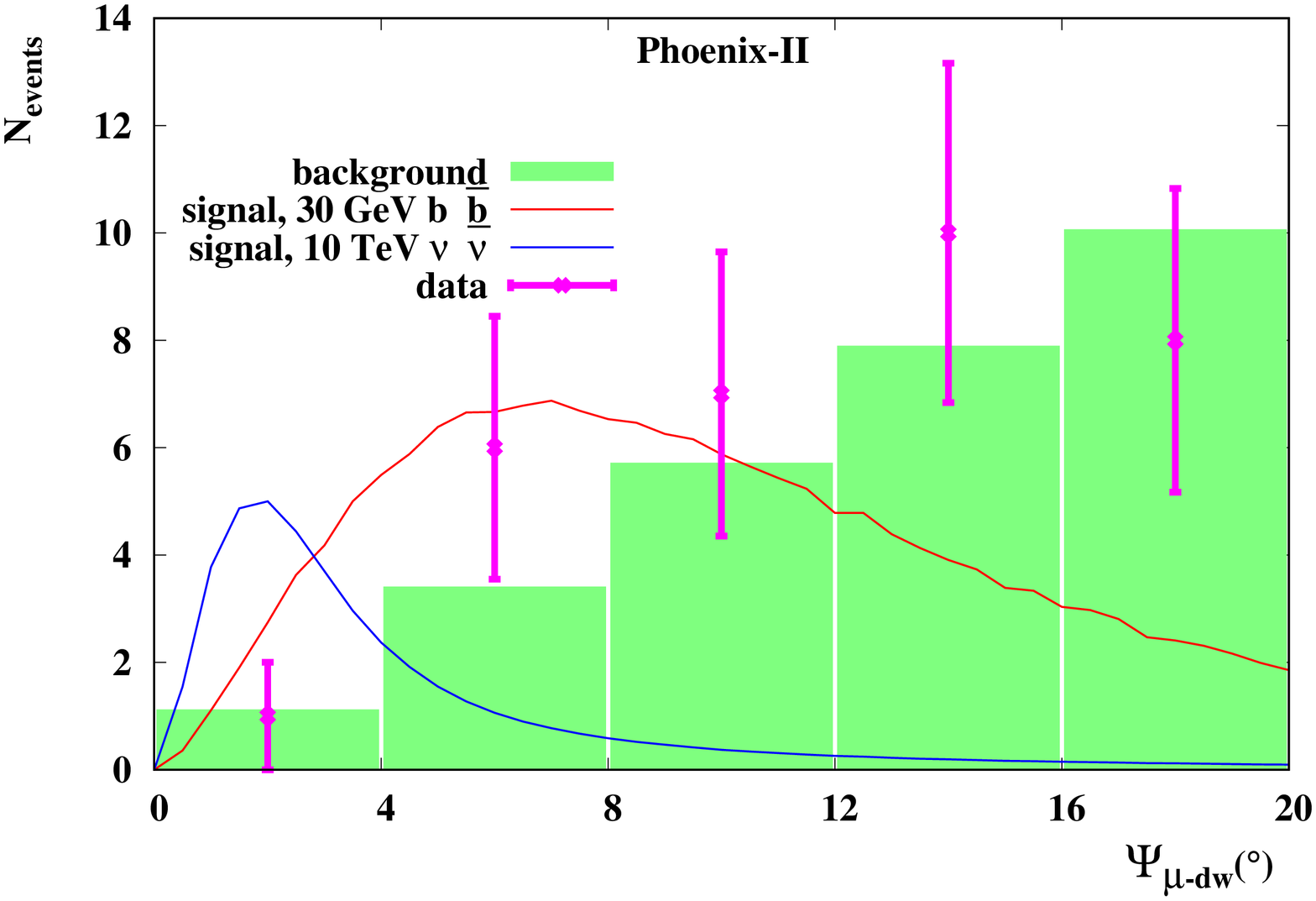} 
    \end{tabular}
  \end{center}
  \caption{\label{fig2:dwarfs}
    Angular distributions of signal ($b\bar{b}$, $m_{DM}=30$~GeV and
    $\nu\bar{\nu}$, $m_{DM}=10~TeV$) in comparison with the data and
    background events ``Eridanus-3'' (left) and ``Phoenix-2'' (right).}
\end{figure}

We start our consideration with the chosen set of dwarf spheroidal
galaxies. Firstly, we look for an excess of events in the direction of
each individual source in the search region neglecting presence of
other sources of dark matter annihilations. We do not find
any excess in the directions towards any of 22 chosen dwarf
spheroidals as compared to the expected background. To quantify
potential deviations from the background only hypothesis we calculate
test statistics (TS) which is defined as $\lambda(0)$, see
eq.~\eqref{eq:5}. Values of TS for two benchmark annihilation channels
$\nu\bar{\nu}$ with $m_{DM}=10$~TeV and $b\bar{b}$ with
$m_{DM}=30$~GeV are presented in Table~\ref{dwarfs_pars}. These
channels have the most and less narrow angular distribution,
respectively. Difference in the forms of these distributions results
in distinction in the values of TS. Maximal deviation from the background only
hypothesis has been observed for ``Eridanus-3'' source; the largest TSs
reach values about 5  for large mass region which correspond to more
collimated signal distributions. In Fig.~\ref{fig2:dwarfs}
we present distribution of observed events in comparison with
the background and signal distribution for two selected sources,
``Eridanus-3'' (left) and ``Phoenix-2'' (right). In the former case
we observe an excess of events at small angles with respect to the
source which resulted in an increase of TS for channels with large
masses of dark matter particles. In the case of ``Phoenix-2'', the
data prefer more wide signal angular distribution because there is an
increase of events at moderate values of $\psi$, in the region 
$5-15^\circ$. For the case of the LMC the test statistics appears to be
zero for all chosen annihilation channels due to found deficit in
observed events as compared to background.

For the following part of the analysis out of the whole set we select
5 dwarf spheroidals, such that they have the largest values of
$J_a$-factors and good visibility on the one hand and expected signal
regions for every galaxy in this set do not intersect with each
other. The selected set contains ``Sculptor'', ``Coma Berenices'',
``Segue-1'', ``Reticulum-2'', ``Tucana-2''. This choice allows us not
only to obtain the  
individual upper limits on annihilation cross section but also to find  
the upper limits from combination of observations of several
independent sources\footnote{The case when the galaxies have small
  angular distance is more involved and requires additional simulation
  of signal angular distribution.}.

\begin{figure}
\begin{center}
\includegraphics[width=0.33\textwidth,angle=-90]{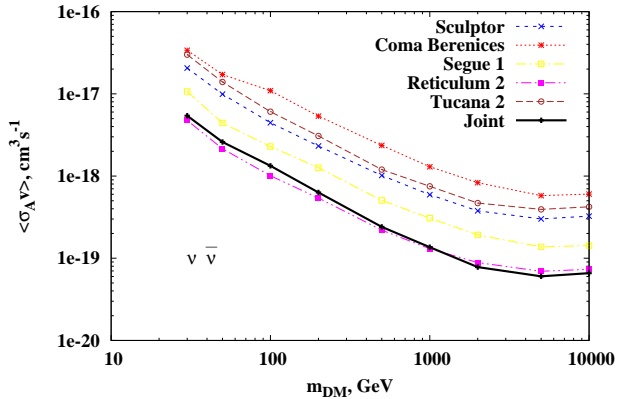}
 \end{center}
\caption{\label{fig:dwarfs} 
  90\% CL upper limits from the NT200 data on dark matter annihilation 
  cross section assuming annihilation to $\nu\bar{\nu}$ for selected
  set of dwarf spheroidal galaxies and from their combination.  }  
\end{figure}

\begin{figure}
\begin{center}
\includegraphics[width=0.33\textwidth,angle=-90]{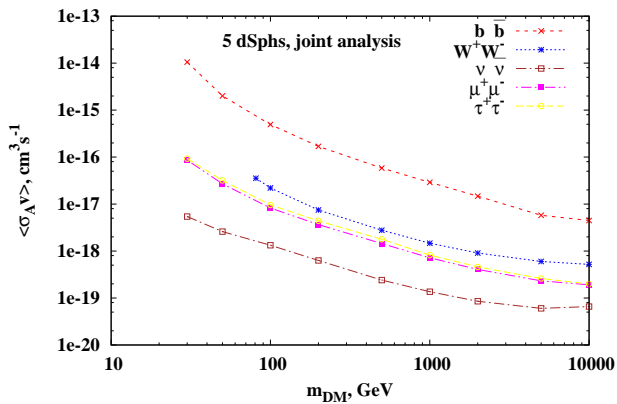}
 \end{center}
\caption{\label{fig:dwarfs_ch} 
  90\% CL upper limits from the NT200 data on dark matter annihilation
  cross section assuming annihilation to $b\bar{b}$, $W^+W^-$,
  $\mu^+\mu^-$, $\tau^+\tau^-$ and $\nu\bar{\nu}$ from joint analysis
  of dwarf spheroidals.  }   
\end{figure}

The individual and joint 90\% upper limits on $\langle\sigma_a
v\rangle$ are obtained by using Feldman-Cousins
approach~\cite{Feldman:1997qc}. In Fig.~\ref{fig:dwarfs} we show the
upper limits 
for 5 chosen dwarfs and the results of combined analysis (marked as
{\it Joint}) for $\nu\bar{\nu}$ annihilation channel. The strongest
individual upper limit comes from observation of ``Reticulum-2''
direction. It is determined by the largest annihilation $J_a$-factor
and the position of this source on the sky which allows for 100\%
visibility from the NT200 position. Although ``Segue-1'' has
similar value of  
$J$-factor with smaller astrophysical uncertainty than those of
Reticulum-2, this source is visible only about 32\% of the whole
lifetime.
The limits from directions towards ``Sculptor'', ``Coma Berenices''
and ``Tucana-2'' are even weaker. The result of combined
analysis is 
determined mainly by the strongest bound from
``Reticulum-2''. Similar 
picture is obtained for other annihilation channels. We compare the
upper limits from combined analysis of dwarfs from different channels
in Fig.~\ref{fig:dwarfs_ch}. 
The strongest upper limit on annihilation cross section from this
search reach values about $6\cdot10^{-20}$~cm$^3/$s for monochromatic
neutrino annihilation channel. Flattening of the results for all
channels (especially for $\nu\bar{\nu}$) for very large masses 
of dark matter particles comes from softer neutrino energy spectrum
which results from enhancement due to electroweak
corrections~\cite{Cirelli:2010xx}.  

\begin{figure}
\begin{center}
\includegraphics[width=0.33\textwidth,angle=-90]{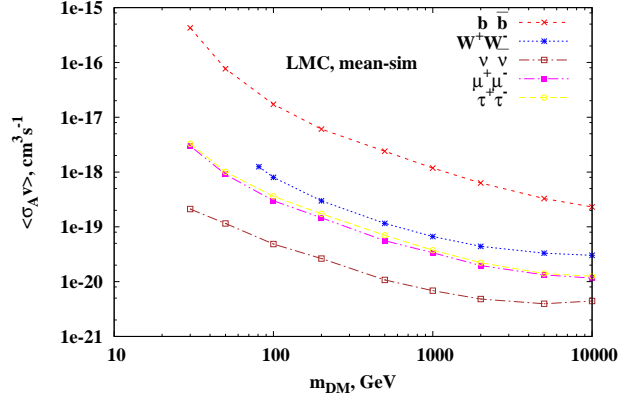}
 \end{center}
\caption{\label{fig:LMC_ch} 
  90\% CL upper limits from the NT200 data on dark matter annihilation
  cross section assuming annihilation to $b\bar{b}$, $W^+W^-$,
  $\mu^+\mu^-$, $\tau^+\tau^-$ and $\nu\bar{\nu}$ from analysis of the
  LMC direction.  }    
\end{figure}

Next, we consider Large Magellanic Cloud as the source of products of
dark matter annihilations.  As we
  discussed above we use {\it
  sim-mean} profile as a default for the LMC and the profiles marked {\it
  sim-min} and {\it sim-max} to estimate astrophysical uncertainty.
In Fig.~\ref{fig:LMC_ch}
we show 90\% CL upper limits on dark matter annihilation cross
section for different annihilation channels. The most stringent bounds
are obtained for $\nu\bar{\nu}$ and reach values about
$7\cdot10^{-21}$cm$^3/$s. 

\begin{figure}
\begin{center}
\includegraphics[width=0.33\textwidth,angle=-90]{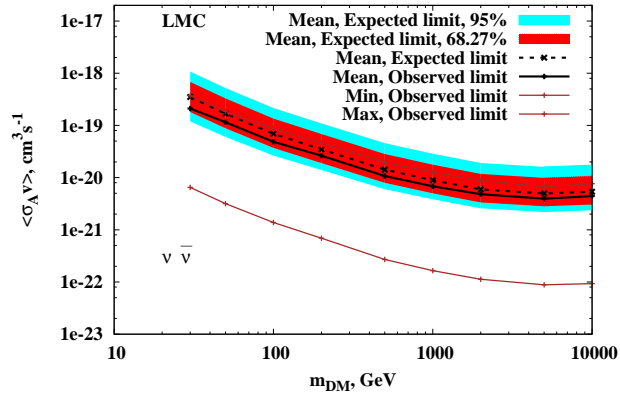}
 \end{center}
\caption{\label{fig:LMC} 
  90\% CL upper limits from the NT200 data assuming different dark
  matter density profiles for LMC (solid lines) and sensitivity
  (dashed line) on dark matter   annihilation cross section assuming
  annihilation to   $\nu\bar{\nu}$. Colored bands represent 68\% (red)
  and 95\% (blue) quantiles.  }    
\end{figure}

Using estimated background we run a set of pseudoexperiments to find
the sensitivity of the NT200 to the neutrino signal from dark matter
annihilations in the LMC. We show this sensitivity for
$\nu\bar{\nu}$ annihilation channel in Fig.~\ref{fig:LMC} (at 1- and
2-$\sigma$ level)   
along with 68\% (red) and 95\% (blue) quantiles in comparison with the 
obtained 90\% CL upper limit shown by the black solid line assuming
{\it sim-mean} profile. Also in this Figure we show 90\% CL upper
limits for this annihilation channel obtained with the other dark
matter density profiles, {\it sim-min} and {\it sim-max}, which can be
viewed as an estimate of astrophysical systematics related to this
source. We see that with ``cuspy'' {\it sim-max} profile the upper
bounds are  improved by almost two orders of magnitude.  

\begin{figure}
\begin{center}
\includegraphics[width=0.33\textwidth,angle=-90]{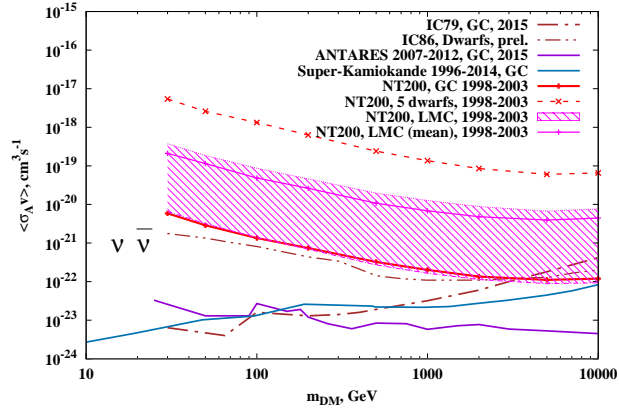}
 \end{center}
\caption{\label{fig:nu_comp} 
  90\% CL upper limits on dark matter annihilation cross
  section assuming annihilation to $\nu\bar{\nu}$ obtained with the NT200
dataset (magenta line and shaded area) in this study and in the GC
(red line) analysis~\cite{NT200GC:2015} in comparison with results
from 
ANTARES~\cite{Adrian-Martinez:2015wey}, IceCube~\cite{Aartsen:2015xej}
and Super-Kamiokande~\cite{Frankiewicz:2015zma}. }
\end{figure}

\begin{figure}
\begin{center}
\includegraphics[width=0.33\textwidth,angle=-90]{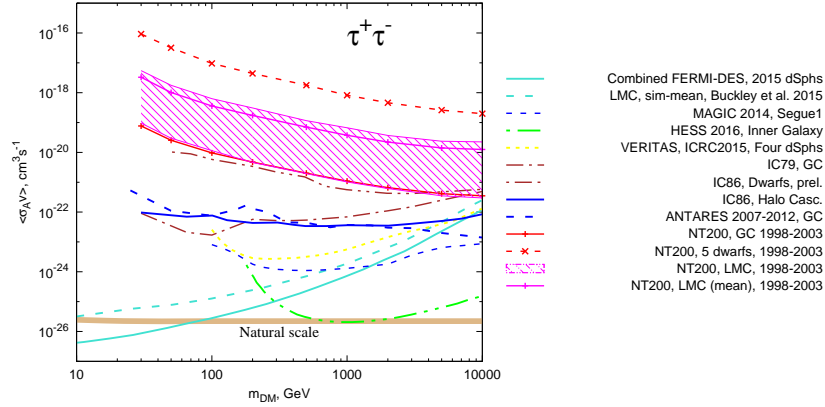}
 \end{center}
\caption{\label{fig:tau_comp} 
  90\% CL upper limits on dark matter annihilation cross
  section assuming annihilation to $\tau^+\tau^-$ in comparison with
  other experiments.  }    
\end{figure}

In Fig.~\ref{fig:nu_comp} 
we present a comparison of upper limits obtained by different neutrino 
experiments from their searches for the dark matter annihilation
signal in comparison with the NT200 results. There are shown the limits
from IceCube (GC~\cite{Aartsen:2015xej} and preliminary
results from joint analysis of dwarf galaxies~\cite{Aartsen:2015bwa}), 
ANTARES (GC~\cite{Adrian-Martinez:2015wey}),
Super-Kamiokande (GC~\cite{Frankiewicz:2015zma}) as well as NT200
limit from the analysis of GC~\cite{NT200GC:2015}. 
In Fig.~\ref{fig:tau_comp}
we compare the 90\% CL upper limits on annihilation cross section
for $\tau^+\tau^-$ annihilation channel obtained by different
experiments. These experiments include the
FERMI~\cite{Drlica-Wagner:2015xua} (dwarf galaxies, DES),
VERITAS~\cite{Zitzer:2015eqa} (four dwarf galaxies), 
MAGIC~\cite{Aleksic:2013xea}  (Segue~1),
HESS~\cite{::2016jja} (inner Galactic halo),
IceCube (Milky Way~\cite{Aartsen:2016pfc}, GC~\cite{Aartsen:2015xej}
and  preliminary results for dwarf galaxies~\cite{Aartsen:2015bwa}), 
ANTARES~\cite{Adrian-Martinez:2015wey} (GC).  Light brown line shows
the thermal relic annihilation cross section from
Ref.~\cite{Steigman:2012nb}.  From the present analysis we see that
for Baikal experiment the LMC direction is more sensitive (even with
astrophysical systematics) to dark matter annihilation signal as
compared to that of dwarf spheroidal galaxies. 

\section {Conclusions}
To summarize, we presented our new results in indirect search for dark
matter signal from distant astrophysical sources, namely the Large
Magellanic Cloud and dwarfs spheroidal galaxies, with neutrino events
of the NT200 neutrino telescope in Lake Baikal. No significant excess in
these directions has been found. We obtained the upper limits at 90\%
CL on annihilation cross sections for different annihilation channels
and masses of dark matter particles in the range from 30~GeV to
10~TeV.  In the present study, the strongest bound on dark matter
annihilation cross section has been found is about $6\cdot 10^{-20}$~cm$^3/$s
for combination of results from dwarfs and about $7\cdot
10^{-21}$~cm$^3/$s for the LMC. We expect considerable
  improvement of these results with data coming from Baikal-GVD
  experiment~\cite{Avrorin:2015skm}.

The work of S.V.~Demidov and O.V.~Suvorova was supported by the RSCF
grant 14-12-01430.

\end{onehalfspace}
\end{document}